\documentclass[onecolumn,prd,aps,nofootinbib]{revtex4}
\usepackage{graphicx}
\usepackage{dcolumn}
\usepackage{bm}
\usepackage{amsmath}
\usepackage{color}
\usepackage{epsfig}
\usepackage{graphicx}
\usepackage{hyperref}
\usepackage{url}

\def\ie{{\frenchspacing\it i.e.}}

\def\be{\begin{equation}}
\def\ee{\end{equation}}
\def\bea{\begin{eqnarray}}
\def\eea{\end{eqnarray}}

\begin{document}

\title{On Dark Energy Isocurvature Perturbation}

\author{Jie Liu${^a}$}
\email{liujie@ihep.ac.cn}
\author{Mingzhe Li$^{b,d}$}
\email{limz@nju.edu.cn}
\author{Xinmin Zhang$^{a,c}$}
\email{xmzhang@ihep.ac.cn} \affiliation{${}^a$Institute of High
Energy Physics, Chinese Academy of Science, P.O. Box 918-4,
Beijing 100049, P. R. China} \affiliation{${}^b$ Department of
Physics, Nanjing University, Nanjing 210093, P.R.China}
\affiliation{${}^c$Theoretical Physics Center for Science
Facilities (TPCSF), Chinese Academy of Science, Beijing 100049,
P.R.China} \affiliation{${}^d$Joint Center for Particle, Nuclear
Physics and Cosmology, Nanjing University - Purple Mountain
Observatory, Nanjing 210093, P. R. China}

\begin{abstract}
Determining the equation of state of  dark energy with
astronomical observations is  crucially important to understand
the nature of dark energy. In performing a likelihood analysis
of the data, especially of the cosmic microwave background
 and large scale structure data the dark energy perturbations
have to be taken into account both for theoretical consistency
and for numerical accuracy. Usually, one assumes in the global fitting analysis
 that the dark energy perturbations are adiabatic. In this paper, we study the dark energy isocurvature
perturbation analytically and discuss its implications for the cosmic microwave background radiation and
large scale structure. Furthermore, with the current astronomical observational
data and by employing Markov Chain Monte Carlo method, we perform a global analysis of
cosmological parameters assuming general initial conditions for the dark energy perturbations.
The results show that the dark energy isocurvature perturbations
are very weakly constrained and that  purely adiabatic initial conditions are consistent with the data.
\end{abstract}

\maketitle

\section{introduction}
\label{intro}

Since the discovery of the accelerated expansion of the universe by
observations of distant Type Ia supernovae (SNIa) in 1998
\cite{Riess:1998cb, Perlmutter:1998np},  dark energy has become a
hot topic in physics and astronomy. So far a lot of models have been
proposed in the literature. In general these models can be
classified according to the equation of state (EoS) $w_e$ of the
dark energy, defined as the ratio of its pressure to energy density.
The simplest  assumption is to consider dark energy with constant
$w_e$, and more specifically to assume a cosmological constant whose
EoS is equals to $-1$. Although this scenario is consistent with
observations\cite{Jassal:2006gf}, it  suffers from the well-known
fine-tuning and coincidence problems\cite{weinberg,CCproblem}.
Alternatively,  dynamical dark energy models, such as quintessence
\cite{Ratra:1987rm, Wetterich:1987fm,Caldwell:1997ii}, phantom
\cite{Caldwell:1999ew}, k-essence \cite{ArmendarizPicon:2000dh},
quintom \cite{Feng:2004ad,Cai:2009zp} and so on, have a
time-dependent EoS. For quintessence $w_e\geq -1$, while for phantom
$w_e\leq -1$. But for quintom models, the  EoS crosses the boundary
set by  $w_e= -1$. To investigate dark energy without making use of
specific field models, one often parameterizes the  EoS of dark
energy as\cite{Chevallier:2000qy} \be w_e(a) = w_0+w_a(1-a) \ee
where $a$ is  the scale factor normalized to be 1 at the present
time. We adopt this parametrization in the current work.

Given the fact that a lot of theoretical models exist in the
literature, it is crucially important to use the accumulated high
precision observational data from SNIa, the cosmic microwave
background (CMB)  radiation and large scale structure (LSS) surveys to constrain the
value of $w_e(a)$. Since  dynamical dark energy should
fluctuate in  space as described by  the conservation of the
energy-momentum tensor, in a global analysis with a general time
evolving EoS one should take into account the dark energy perturbations in order
to have a consistent procedure. This is particularly important  when fitting
cosmological parameters to the data of CMB
and LSS. Simply switching off the dark energy perturbation is not
theoretically correct and will lead to biased results. Numerically
it has been shown that the results obtained are quite different
between the two cases with and without the dark energy
perturbations \cite{Bean:2003fb,  Mukherjee:2003rc, Zhao:2005vj, Xia:2005ge,
Weller:2003hw,Xia:2007km, Komatsu:2008hk,Yeche:2005wn}.

With the generally parameterized EoS, there inevitably exists a singularity when $w_e = -1$.
When $w_e$ crosses this critical point, the dark energy perturbations will diverge
 \cite{Feng:2004ad,Vikman:2004dc, Hu:2004kh,Caldwell:2005ai}. It has been shown that in the
context of general relativity, it is impossible to
obtain a background which crosses the ``cosmological constant
boundary" with only a single scalar field or a single perfect
fluid. In fact, this is the reason why the quintom scenario of dark energy needs to introduce extra degrees of freedom
\cite{Feng:2004ad,Vikman:2004dc, Hu:2004kh, Caldwell:2005ai,
Zhao:2005vj, Li:2005fm, xiaofei, Lim:2010yk,Deffayet:2010qz} \footnote{For a consistent and
complete proof of the no-go theorem, please see,
\cite{Xia:2007km}.}.
In order to handle the perturbation when $w_e$ crosses $-1$,
we proposed a method
to deal with the dark energy perturbations during
the crossing of the boundary $w_e= -1$ in Ref. \cite{Zhao:2005vj}.
According to this method, the energy and
momentum density perturbations of dark energy are treated as
constant during the small interval around the critical point
$w_e=-1$. This method is justified in Ref. \cite{Li:2010hm} from
the viewpoint of general relativistic matching conditions.

Another issue concerned with the dark energy perturbations is the
question of initial condition. Generally, there are two types of
initial conditions for the perturbations: adiabatic and
isocurvature. In Refs. \cite{Zhao:2005vj,Xia:2005ge,
Xia:2007km,Li:2010hm} for the global fitting of the dark energy EoS
to the observational data, it was assumed that the perturbations
were purely adiabatic.  In this paper we will study more general
initial conditions which admit dark energy isocurvature
perturbations and discuss the implications for CMB temperature and
polarization power spectra and the LSS  matter power spectrum. In
the literature, baryon and dark matter isocurvature perturbations
have been extensively discussed and tight constraints on these are
obtained. There have also been studies of the dark energy
isocurvature perturbations which, however, are usually limited in
the framework of quintessence models \cite{Bucher:1999re,
Kawasaki:2001bq, Malquarti:2002iu, Moroipq, Gordon:2004ez}. In these
studies, it has been shown that the quintessence isocurvature
perturbations could lead to the suppression of the CMB quadrupole
via the anti-correlation between the adiabatic and the isocurvature
modes\cite{Moroipq, Gordon:2004ez,Li:2010ac,Enqvist:2000hp}. In this
paper, working with the parameterized EoS, we consider both
adiabatic and isocurvature initial conditions and the correlation
between them in the likelihood data fitting analysis. We discuss the
current constraints on the cosmological parameters, with result when
admitting the possible existence of dark energy isocurvature modes.
Our paper is organized as follows: in section II, we briefly review
the theory of perturbations; In section III, we  analytically study
in detail the dark energy isocurvature perturbations; In section IV
we study effects of the dark energy isocurvature perturbations on
CMB and LSS,  and  we present the current constraints on them in
Section V; Section VI is our summary.

\section{Adiabatic and Isocurvature perturbations}

We consider a spatially flat Friedmann-Robertson-Walker universe as
the background. The metric of the perturbed spacetime in the
conformal Newtonian gauge reads, \be ds^2 =
a(\eta)^2[(1+2\Phi)d\eta^2-(1-2\Phi)\delta_{ij}dx^idx^j]~, \ee where
we have implicitly assumed that the shear perturbations can be
neglected and the metric perturbations are fully described by one
relativistic potential $\Phi$. In the matter sector, the
perturbations are expressed by the perturbed energy-momentum tensor
which is gauge dependent. However, for the discussions of
perturbations on large scales it is more convenient to use gauge
invariant variables constructed by combining the energy-momentum
perturbations with the metric perturbations. In this paper, we use
the following gauge-independent variables for each species \bea
\zeta_{\alpha}&=& \frac{\delta_{\alpha}}{3(1+w_{\alpha})}-\Phi~,\nonumber\\
\Delta_{\alpha} &=&
\frac{\rho_{\alpha}\delta_{\alpha}}{3}+\frac{\mathcal{H}}{k^2}(\rho_{\alpha}+p_{\alpha})\theta_{\alpha}~,
\eea where $\delta_{\alpha} \equiv
\delta\rho_{\alpha}/\rho_{\alpha}$ is the density contrast,
$\theta_{\alpha} \equiv ik^i\delta
T^0_{~i\alpha}/(\rho_{\alpha}+p_{\alpha})$ is the corresponding
momentum density perturbation,  and the conformal Hubble parameter
is defined by $\mathcal{H}=a'/a$ with the prime denoting the
derivative with respect to conformal time. $\zeta_{\alpha}$ is a
comoving curvature perturbation, and as we will see later in this
paper $\Delta_{\alpha}$ may be called an ``effective" density
perturbation. The conservation of the energy-momentum tensor at the
linear order gives the equations governing the evolutions of
$\zeta_{\alpha}$ and $\Delta_{\alpha}$: \bea &
&\zeta_{\alpha}'+3\mathcal{H}(c_{s\alpha}^2-c_{a\alpha}^2)\frac{\Delta_{\alpha}}{\rho_{\alpha}+p_{\alpha}}+
\frac{k^2}{3\mathcal{H}}(\frac{\Delta_{\alpha}}{\rho_{\alpha}+p_{\alpha}}-\zeta_{\alpha})=\frac{k^2}{3\mathcal{H}}\Phi~,\label{zeta}\\
&
&\Delta_{\alpha}'+(4\mathcal{H}-\frac{\mathcal{H}'}{\mathcal{H}}+\frac{k^2}{3\mathcal{H}})
\Delta_{\alpha}-(\mathcal{H}-\frac{\mathcal{H}'}{\mathcal{H}}+\frac{k^2}{3\mathcal{H}})(\rho_{\alpha}+p_{\alpha})\zeta_{\alpha}
=(\rho_{\alpha}+p_{\alpha})[\Phi'+
(2\mathcal{H}-\frac{\mathcal{H}'}{\mathcal{H}}+\frac{k^2}{3\mathcal{H}})\Phi]\label{Theta}~.
\eea
In the above equations, $c_{s\alpha}$ is the sound speed
defined in the comoving frame of the fluid while the so-called
adiabatic sound speed, $c_{a\alpha}$,  is defined as $
c_{a\alpha}^2 \equiv p_{\alpha}'/\rho_{\alpha}' =
w_{\alpha}-w_{\alpha}'/[3 \mathcal{H}(1+w_{\alpha}) ]$. For a
perfect fluid $c_{s\alpha}=c_{a\alpha}$, and for a canonical
scalar field $c_{s\alpha}=1$.

To close the system, we also  need the Poisson equation,
\be\label{Poisson} \frac{k^2}{a^2}\Phi=-12\pi
G\sum_{\alpha}\Delta_{\alpha}~, \ee which can be obtained from the
perturbed Einstein equations. Comparing this equation with the
Poisson equation in the Newtonian gravity we can see that
$\Delta_{\alpha}$ may be called an effective density perturbation.
On super horizon scales $k\eta\ll 1$, the terms proportional to
$k^2$ in Eqs. (\ref{zeta}), (\ref{Theta}) and (\ref{Poisson}) can
be dropped, and these equations become \bea &
&\zeta_{\alpha}'+3\mathcal{H}(c_{s\alpha}^2-c_{a\alpha}^2)\frac{\Delta_{\alpha}}{\rho_{\alpha}+p_{\alpha}}
= 0~,\label{zeta2}\\
& &
\Delta_{\alpha}'+(4\mathcal{H}-\frac{\mathcal{H}'}{\mathcal{H}})
\Delta_{\alpha}-(\mathcal{H}-\frac{\mathcal{H}'}{\mathcal{H}})(\rho_{\alpha}+p_{\alpha})\zeta_{\alpha}
= (\rho_{\alpha}+p_{\alpha})[\Phi'+
(2\mathcal{H}-\frac{\mathcal{H}'}{\mathcal{H}})\Phi]\label{Theta2}~,\\
& &\sum_{\alpha}\Delta_{\alpha}= 0\label{Poisson2}~. \eea
From
equation (\ref{zeta2}), we see that for a perfect fluid
$\zeta_{\alpha}$ is conserved on large scales.

To solve the set of perturbation equations, we need to specify the
initial conditions. Usually these initial conditions are set at
the time deep inside the radiation dominated era when the scales
corresponding to  observations today were far outside the
horizon. Inflation provides a natural mechanism to generate these
initial perturbations. They originate from quantum vacuum
fluctuations and become classical perturbations when their
corresponding length scales leave the horizon during
inflation. In the post-inflation epoch, these primordial
perturbations re-enter the horizon and interact with matter
to cause the CMB anisotropies and structure formation. Hence it is
important to inspect the evolutions of the perturbations on
super-horizon scales until the scales re-enter the horizon.
For this purpose we  only need to consider the equations
(\ref{zeta2}), (\ref{Theta2}) and (\ref{Poisson2}). There are two
types of solutions of these equations, called adiabatic and
isocurvature (or entropy) modes. For adiabatic perturbation all
comoving curvature perturbations $\zeta_{\alpha}$ are the same as
that of radiation $\zeta_r$. The total comoving curvature
perturbation is also equal to $\zeta_r$, \be
\zeta=\frac{1}{\rho+p}\sum_{\alpha}(\rho_{\alpha}+p_{\alpha})\zeta_{\alpha}=\zeta_r~,
\ee which is constant since $\zeta_r'=0$. Thus, for adiabatic
perturbation, the picture is simple: the primordial perturbations
$\zeta$ are frozen while they are outside the horizon. With Eq.
(\ref{Poisson2}), the sum of Eq. (\ref{Theta2}) over all species
gives \be\label{potential}
(\frac{\mathcal{H}'}{\mathcal{H}}-\mathcal{H})\zeta=\Phi'+
(2\mathcal{H}-\frac{\mathcal{H}'}{\mathcal{H}})\Phi~. \ee
Integration of this equation gives \be\label{adiabatic} \Phi^{\rm
adi}=C\frac{\mathcal{H}}{a^2}-\zeta_r(1-\frac{\mathcal{H}}{a^2}\int\frac{a
da}{\mathcal{H}(a)})~, \ee where $C$ is a constant. The first term
on the right hand side decays in the expanding universe and can be
neglected.

However, if one of the comoving curvature perturbations
$\zeta_{\alpha}$ is not equal to $\zeta_r$, the density perturbation
have an isocurvature mode.
the isocurvature perturbation of species $\alpha$ is defined by
\be
S_{\alpha}\equiv 3(\zeta_{\alpha}-\zeta_r)=\frac{\delta_{\alpha}}{1+w_{\alpha}}-\frac{3}{4}\delta_r~.
\ee
In principle,  if the universe contains $N$ components, there should be
at most $N-1$ isocurvature density perturbations.
With the presence of isocurvature modes,
the equation (\ref{potential}) relating the potential and the
total comoving curvature perturbation is still valid. However
in this case $\zeta\neq \zeta_r$. If only one species $\alpha$
has isocurvature perturbation, $\zeta$ is
\be\label{iso1}
\zeta=\zeta_r+\frac{\rho_{\alpha}+p_{\alpha}}{3(\rho+p)}S_{\alpha}~.
\ee
It is not conserved on large scales.
We may define $\xi_{\alpha}\equiv (\rho_{\alpha}+p_{\alpha})S_{\alpha}/3$, so that
\be\label{iso2}
\zeta=\zeta_r+\frac{\xi_{\alpha}}{\rho+p}~,
\ee
and Eqs. (\ref{zeta2}) and (\ref{Theta2}) become
\bea\label{evolution}
& &\xi_{\alpha}'+3\mathcal{H}(1+c_{a\alpha}^2)\xi_{\alpha}+3\mathcal{H}(c_{s\alpha}^2-c_{a\alpha}^2)\Delta_{\alpha}=0\nonumber~,\\
&
&\Delta_{\alpha}'+(4\mathcal{H}-\frac{\mathcal{H}'}{\mathcal{H}})\Delta_{\alpha}+(\frac{\mathcal{H}'}{\mathcal{H}}-\mathcal{H})
(1-\frac{\rho_{\alpha}+p_{\alpha}}{\rho+p})\xi_{\alpha}=0~. \eea
The above equations describe how the isocurvature perturbations
evolve on large scales and Eq. (\ref{iso1}) or (\ref{iso2})
characterizes the contribution of isocurvature perturbations to the
total comoving curvature perturbation. The potential can be solved
by integrating Eq. (\ref{potential}) and we obtain \be\label{iso3}
\Phi=\Phi^{\rm adi}-4\pi G\frac{\mathcal{H}}{a^2}\int
\xi_{\alpha}\frac{a^3}{\mathcal{H}(a)^3}da~, \ee
where $\Phi^{\rm
adi}$ is the contribution of the adiabatic mode given in Eq.
(\ref{adiabatic}),  and the last term is the contribution from the
isocurvature perturbation. To get it we have used the equation
$\mathcal{H}'-\mathcal{H}^2=-4\pi Ga^2(\rho+p)$.
This equation shows explicitly that both adiabatic and isocurvature perturbations
are able to generate the metric perturbation.

If the species $\alpha$ is subdominant
$(\rho_{\alpha}+p_{\alpha})/(\rho+p)\rightarrow 0$, its perturbation
has a negligible contribution to the metric perturbation and the
potential $\Phi\simeq \Phi^{\rm adi}$. Then the evolution equations
(\ref{evolution}) of isocurvature perturbations become
\bea\label{evolution2}
& &\xi_{\alpha}'+3\mathcal{H}(1+c_{a\alpha}^2)\xi_{\alpha}+3\mathcal{H}(c_{s\alpha}^2-c_{a\alpha}^2)\Delta_{\alpha}=0\nonumber~,\\
&
&\Delta_{\alpha}'+(4\mathcal{H}-\frac{\mathcal{H}'}{\mathcal{H}})\Delta_{\alpha}+(\frac{\mathcal{H}'}{\mathcal{H}}-\mathcal{H})
\xi_{\alpha}=0~. \eea These two equations are the basis for the
discussion of dark energy isocurvature perturbations in the
radiation and matter dominated eras in the next section. We can see
from Eq. (\ref{iso2}) that the contribution of dark energy
isocurvature perturbations relies on the ratio $\xi_e/(\rho+p)$
compared with $\zeta_r$, where the subscript $e$ represents dark
energy. Because $\zeta_r$ is conserved, qualitatively the effect of
the isocurvature depends on whether $\xi_e/(\rho+p)$ grows or decays
with time. When the density of dark energy becomes significant at
late time, its isocurvature perturbations could make as important
contribution to the metric perturbation and we should use the
equations (\ref{evolution}) to investigate its evolution.

\section{Dark Energy Isocurvature Perturbation}

In this section we will study the dark energy isocurvature
perturbations during the radiation and matter dominated epochs.
For simplicity we assume that the perturbations
of baryons, dark matter, neutrinos and so on are adiabatic.
The dark energy was subdominant in the early universe and
it only starts to dominate the universe at low redshifts. So we can use the equations (\ref{evolution2}) to study the
dark energy isocurvature perturbation on super-horizon scales.

First of all, we will discuss the behavior of dark energy
isocurvature perturbations for some specific dark energy models.

\subsection{Single Fluid}

If dark energy is a perfect fluid with one component, its sound speed in
its comoving frame is equal to the adiabatic sound speed,
$c_{se}^2=c_{ae}^2$. It cannot be negative otherwise its
perturbation would be unstable on small scales. From the first equation
of (\ref{evolution2}), we have \be
\xi_{e}'+3\mathcal{H}(1+c_{ae}^2)\xi_{e}=0~. \ee This equation
implies \be \xi_e\propto \rho_e+p_e~. \ee So, the contribution $\xi_e/(\rho+p)$  of
dark energy isocurvature perturbation scales as
$(\rho_e+p_e)/(\rho+p)$. If $c_{ae}^2$ changes slowly, we can
treat it as a constant. In this case, $\xi_e/(\rho+p)$ will not be
damped if $c_{ae}^2\leq 1/3$ in the radiation epoch. However,
$\xi_e/(\rho+p)$ will decay in the matter dominated era unless
$c_{ae}^2= 0$.

\subsection{Multiple fluids}

If the dark energy contains multiple perfect fluids, then for each component we have $\xi_{ei}=C_i (\rho_{ei}+p_{ei})$ and
\bea
\xi_e&=&\sum_i \xi_{ei}=\sum_i C_i (\rho_{ei}+p_{ei})~,\nonumber\\
\frac{\xi_e}{\rho+p}&=&\frac{\sum_i C_i
(\rho_{ei}+p_{ei})}{\rho+p}~, \eea where $C_i$ are constants. If
all the constants $C_i$ are the same, then there are no
internal isocurvature perturbations among the components of dark
energy, $\xi_e/(\rho+p)$ scales as $(\rho_e+p_e)/(\rho+p)$ like
in the case of  single fluid. Otherwise, we should study the evolution of the
isocurvature perturbation for each component individually.

\subsection{Single Field}

There are many proposals for dark energy models based on scalar
fields. Simple dark energy models like quintessence, phantom
and k-essence can be constructed from one scalar field. For
quintessence or phantom dark energy, the sound speed is $c_{se}^2=1$.
For k-essence, $c_{se}^2$ can have any nonnegative value.
In general the behavior of scalar field dark energy models
cannot be solved analytically. But in some cases,
$c_{se}^2$ and the equation of state $w_{e}$ change very slowly
 and can be treated as constants, and consequently the analysis becomes simpler.
 In this case the equations (\ref{evolution2}) reduce to the  following second
 order differential equation
 for $\xi_e$,
\be\label{evolution3}
\xi_e''+[(7+3w_e)\mathcal{H}-2\frac{\mathcal{H}'}{\mathcal{H}}]\xi_e'+3[(4+3w_e+c_{se}^2)\mathcal{H}^2-(1+c_{se}^2)\mathcal{H}']\xi_e=0~.
\ee In the radiation dominated era, $a\propto \eta$ and
$\mathcal{H}=1/\eta$, and the solution of Eq. (\ref{evolution3}) is
\be \xi_e=\eta^{-(8+3w_e)/2}[C_1
\eta^{\frac{1}{2}\sqrt{(3w_e+2)^2-24
c_{se}^2}}+C_2\eta^{-\frac{1}{2}\sqrt{(3w_e+2)^2-24 c_{se}^2}}]~.
\ee The contributions of dark energy isocurvature perturbations to
the metric perturbation scales as \be
\frac{\xi_e}{\rho_r+p_r}\propto \eta^{-3w_e/2}[C_1
\eta^{\frac{1}{2}\sqrt{(3w_e+2)^2-24
c_{se}^2}}+C_2\eta^{-\frac{1}{2}\sqrt{(3w_e+2)^2-24 c_{se}^2}}]~.
\ee For quintessence or phantom dark energy, $c_{se}^2=1$ and $
w_e\leq 1$, we can see from the  above equation that the effect of
the isocurvature perturbations of any model with positive $w_e$
decays with the expansion of the universe. If $w_e=0$,
$\xi_e/(\rho_r+p_r)$ oscillates with constant amplitude. For
negative equation of state, the effect of the isocurvature
perturbations grows with time. An interesting case is that when the
scalar field is almost frozen during the radiation epoch, i.e.,
$w_e\simeq -1$, $\xi_e/(\rho_r+p_r)$ grows as $a^{3/2}$ even though
$\xi_e$ and $S_e$ decrease with time.

For k-essence dark energy, $c_{se}^2$ may not be equal to one, but
it should be nonnegative. In this case, for the models with
smaller sound speed, the contribution of its isocurvature
perturbation is more likely to be growing.

Similarly in the matter dominated era, $a\propto \eta^2$ and $\mathcal{H}=2/\eta$. Thus,
for constant $w_e$ and $c_{se}^2$, one has
\be
\frac{\xi_e}{\rho+p}\simeq \frac{\xi_e}{\rho_m}=\eta^{-(3+6w_e)/2}[C_1 \eta^{\frac{3}{2}\sqrt{(1+2w_e)^2-8c_{se}^2}}+
C_2\eta^{-\frac{3}{2}\sqrt{(1+2w_e)^2-8c_{se}^2}}]~.
\ee
This shows that  the contribution of dark energy isocurvature perturbations
has a similar behavior as  in the case of radiation dominated era as
discussed above. With more negative equation of state or smaller sound speed, this contribution is more likely to grow with time.

\subsection{Multiple fields}

If dark energy contains multi-fields, like in the case of the quintom model, we
should solve  the equations (\ref{evolution2}) for each component.
Here, for simplicity,  we have assumed  that there are
no interactions other than gravity among the internal components
of the dark energy. The total contribution of dark energy
isocurvature perturbations is \be \frac{\xi_e}{\rho+p}=\frac{\sum_i
\xi_{ei}}{\rho+p}~. \ee For example, consider
 the quintom model with two fields,
one quintessence and the other a phantom field, and  assume that
each field has an extremely small mass. Both fields are slowly
rolling in the radiation dominated era and
$\xi_{ei}/(\rho_r+p_r)\propto a^{3/2}$ for $i=1,~2$. Hence we have
\be \frac{\xi_e}{\rho_r+p_r}\propto a^{3/2}~. \ee

\section{The effects of dark energy isocurvature perturbation}

In this section we study the effects of the dark energy isocurvature
perturbations on CMB and LSS observations. We will take the
parametrization $w_{e}(a) = w_0 + w_a(1-a)$ and consider the sound
speed $c_{se}^2$ as an arbitrarily non-negative parameter. Besides,
we also need to parameterize the power spectra of the initial
perturbations. Statistically, both adiabatic perturbation $\zeta_r$
and isocurvature perturbation $S_e$ are treated as random fields as
predicted by inflation theory.To be general, we should consider the
correlation between them. If their statistics are Gaussian, both the
adiabatic and isocurvature fields are fully described by the power
spectra. To characterize a well defined system including both
adiabatic and isocurvature modes, one usually introduce a vector
$\mathcal{X}_i$ with two components, \bea \mathcal{X}_i=
\begin{cases}
\zeta_r &{\rm adiabatic}~, \\
S_e  &{\rm isocurvature}~.
\end{cases}
\eea Then the primordial power spectra $\mathcal{P}_{ij}$ are defined by
\be
\langle\mathcal{X}_i(\textbf{k})\mathcal{X}^\ast_j(\textbf{k}')\rangle=\frac{2\pi^2}{k^3}\mathcal{P}_{ij}(k)\delta(\textbf{k}-\textbf{k}').
\ee

One can parameterize the power spectra as
$\mathcal{P}_{ij}=A_{ij}(\frac{k}{k_0})^{n_{ij}-1}$, where  $A_{ij}$
and $n_{ij}$ are $2-$dimensional matrices which characterize the
amplitudes and spectral indices, respectively. We have
 \bea A_{ij}=\begin{pmatrix}
A_{\rm adi}&\sqrt{A_{\rm adi}A_{\rm iso}}\cos\Delta\\
 \sqrt{A_{\rm adi}A_{\rm iso}}\cos\Delta & A_{\rm iso}
\end{pmatrix}~,
\eea where $\cos\Delta=\frac{A_{\rm adi,iso}}{\sqrt{A_{\rm
adi}A_{\rm iso}}}$ describes the correlation between adiabatic and
isocurvature perturbations \cite{Langlois:1999dw}, and $A_{\rm adi}$
and $A_{\rm iso}$ are the amplitudes of adiabatic and isocurvature
modes respectively. The spectral indices are denoted by $n^{\rm
adi}_s=n_{11}$ and $n^{\rm iso}_s=n_{22}$. For simplicity we assume
that $n^{\rm cor}_s=n_{12}=\frac{n_{11}+n_{22}}{2}$
\cite{KurkiSuonio:2004mn}.

As we see from Eq. (\ref{iso3}), both  adiabatic and isocurvature modes can generate  metric perturbations and therefore
temperature anisotropies.
Symbolically we have
\bea
\frac{\delta T}{T}&=& (\frac{\delta T}{T})_{\rm
adi}+(\frac{\delta T}{T})_{\rm iso}, \eea and hence, the temperature angular spectrum $C_l$ can be
expressed as \be C_l = A_{\rm adi}\hat{C}_l^{\rm adi}+A_{\rm
iso}\hat{C}_l^{\rm iso}+2\sqrt{A_{\rm adi}A_{\rm
iso}}\cos\Delta\hat{C}_l^{\rm adi, iso} \ee
where\be \hat{C}_l^{ij}=\frac{4\pi}{2l+1}\int d\ln k(\frac{k}{k_0})^{n_{ij}-1}\Theta_l^i(k)\Theta_l^j(k) \ee
with $\Theta_l^i$
being the transfer function of photons for the initial condition $i$.
There are similar formulas for the CMB $EE$ and $BB$ polarization spectra and temperature-polarization spectrum $TE$.

The isocurvature perturbations also affects the matter power spectrum $P(k)$ as follows,
 \be P(k)
= A_{\rm adi}\hat{P}^{\rm adi}(k)+A_{\rm iso}\hat{P}^{\rm
iso}(k)+2\sqrt{A_{\rm adi}A_{\rm iso}}\cos\Delta\hat{P}^{\rm adi,
iso}(k)~, \label{pk}\ee where $\hat{P}^{\rm ij}(k)$ can be described
as \be \hat{P}^{ij}(k) = (\frac{k}{k_0})^{n_{ij}-1}T^i(k)T^j(k), \ee
with $T^i(k)$ being the transfer functions of matter perturbation
for initial condition $i$.

\begin{figure}
\includegraphics[scale=0.35]{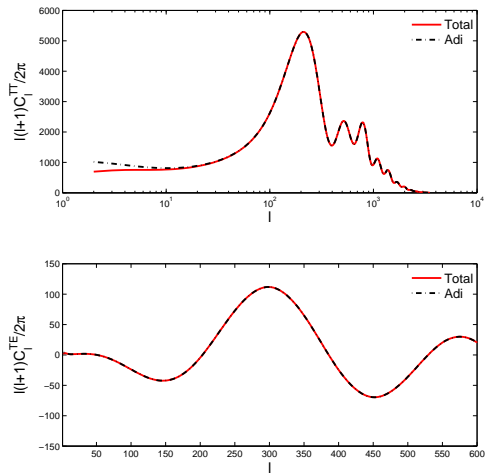}
\caption{Top Panel: The angular power spectrum of CMB. Bottom
Panel: The TE power spectrum of CMB. The red solid lines denote
the spectrum obtained including the contribution of anti-correlated
adiabatic and isocurvature perturbation,
while the black dashed line is obtained by only including
the adiabatic contribution.
}\label{fig:cmb}
\end{figure}

\begin{figure}

\includegraphics[scale=0.35]{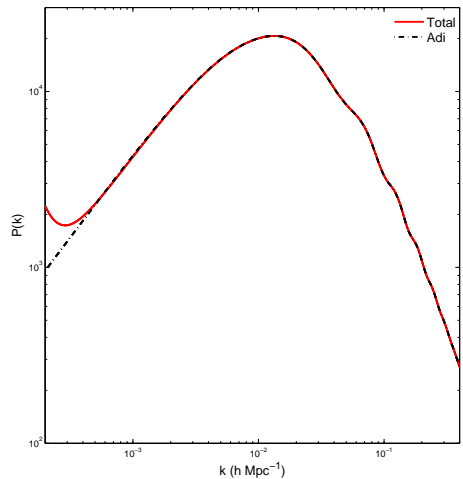}
\caption{The matter power spectrum obtained with the cosmological
parameters chosen to be the same as in Fig.\ref{fig:cmb}. The red line denotes
the total power spectrum and the black dash-dotted line is that
with only the adiabatic component.
}\label{fig:lss}
\end{figure}

In order to show the effects of the isocurvature perturbations on
CMB and LSS observations, we plot in Fig.\ref{fig:cmb} the TT and TE
power spectra of CMB and in Fig.\ref{fig:lss} the matter power
spectrum in the case of fully anti-correlation, $\cos\Delta=-1$. In
the computations,  the fiducial cosmological parameters are chosen
as $w_0 = -1.148$, $w_a = 1.01$, $c_{se}^2 = 0.01$, $A_{\rm
adi}=2.36\times10^{-9}$, $n_s^{\rm adi}=0.95$, $A_{\rm
iso}=1.48\times10^{-8}$, $n_s^{\rm iso}=-1.7$, $\omega_b = 0.02247$,
$\omega_c=0.1135$, $H_0=71.8$km/s/Mpc, where
$\omega_b\equiv\Omega_bh^2$ and $\omega_c\equiv\Omega_bh^2$ denote
the physical baryon and cold dark matter density parameters,
respectively, and $H_0=100h$km/s/Mpc is the current Hubble constant.
The main effects of isocurvature perturbations of dark energy are on
large scales. One can see the suppression in the CMB quadrupole
which is realized by the anti-correlation between isocurvature and
adiabatic perturbations. However, in the matter power spectrum,
there is an increment on large scales with this set of parameters
chosen.

Since the effects of isocurvature perturbations appear mainly at
large scale, it will be difficult to get a tight constraint on it
with current data. This is because we know that in the case of the
CMB, the data on large scale is cosmic variance uncertainty dominated,
while for LSS the largest scale we can observed today is only about $k=0.02 h
{\rm Mpc}^{-1}$ \cite{sdsslrg7}.

\section{observational constraints}
\label{result}
\subsection{Data and  Cosmological Parameters}
We extended the publicly available MCMC package
CosmoMC\footnote{\url{http://cosmologist.info/cosmomc/.}}
\cite{cosmomc} by including  dynamical dark energy and its
perturbations discussed in this paper. We then performed a global
analysis. In the computation of the CMB we have included the WMAP7
temperature and polarization power spectra with the routine for
computing the likelihood supplied by the WMAP
team\footnote{Available at the LAMBDA website:
http://lambda.gsfc.nasa.gov/.}\cite{wmap7}. Furthermore, we include
small scale temperature anisotropies measured by ACBAR \cite{acbar},
CBI \cite{cbi} and Boomerang \cite{boomerang}. The matter power
spectrum measured by observations of luminous red galaxies (LRG)
from SDSS \cite{sdsslrg7}, and the ``Union II'' supernovae dataset
\cite{union2} was also taken into account. Furthermore, we added a
prior on the Hubble constant, $H_0=74.2 \pm 3.6$ km/s/Mpc given by
ref.\cite{HST} as well as a weak Gaussian prior on the baryon
density $\omega_b = 0.022\pm 0.002 ( 1\sigma )$ from Big Bang
Nucleosynthesis\cite{BBN}\footnote{Actually, the chosen of prior may
lead some uncertainty. However, we have checked that the change is
neglectable.}. Simultaneously we also used a cosmic age tophat prior
as $10$ Gyr $< t_0 < 20$ Gyr.

In the numerical calculation we considered the most general
parameter space  \be \mathcal{\bf
P}\equiv\left\{\omega_b,\omega_c,\Theta_s,\tau,w_0,w_a,c_{se}^2,
\cos\Delta, n_s^{\rm adi},n_s^{\rm iso},A_{\rm adi},A_{\rm
iso}\right\}, \label{par} \ee where $\Theta_s\equiv100\frac{r_s}{d_A}$ is the ratio of
the sound horizon to the angular diameter distance at decoupling and
$\tau$ characterizes the optical depth to reionization.

\subsection{Global Fitting Results}

In order to show explicitly the effect of dark energy isocurvature
perturbation, we have done two different kinds of calculations: one
is with pure adiabatic initial condition by making the three
parameters in Eq.(\ref{par}), i.e. $n_s^{\rm iso}, A_{\rm iso}$ and
$\cos\Delta$ vanish and another is the calculation with the full set
of parameters in Eq.(\ref{par}) including the dark energy
isocurvature perturbation (hereafter refer to as ``Mixed''). First
of all we considered the effect of the isocurvature perturbation on
the determination of the dark energy EoS. In Tab.\ref{tab:result} we
present the numerical values of the dark energy EoS.

\begin{figure}[htbp]
\centering
\includegraphics[width=0.5\linewidth]{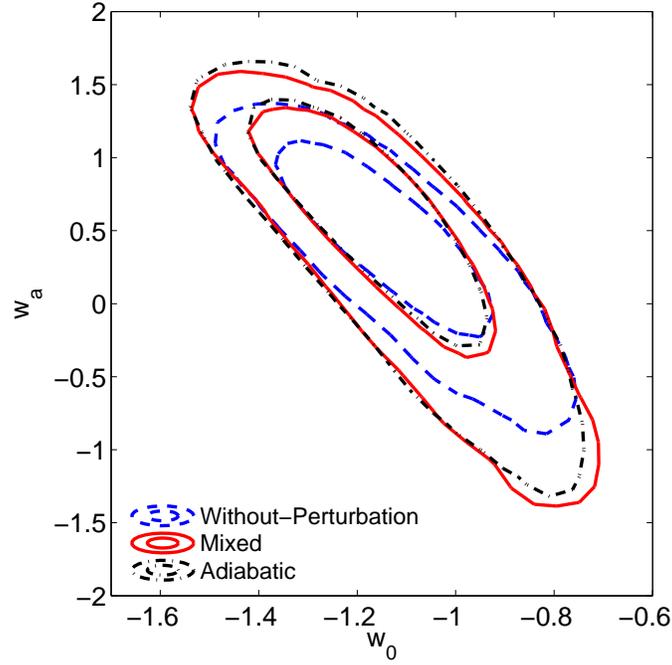}
\caption{The constraints on the dark energy EoS parameter $w_0$ and $w_a$. The red solid lines
are for the mixed case and black dash-dotted line for adiabatic case.The blue dashed lines
stand for the constraints without dark energy perturbation.}\label{fig:w0wa}
\end{figure}

\begin{table}[htbp]
\centering
\begin{tabular}{c|c|c}
\hline\hline
  &  $w_0$ & $w_a$ \\
  \hline
Adiabatic & $-1.143\pm0.160$ & $0.463\pm0.605$ \\
\hline
Mixed & $-1.132^{+0.164}_{-0.158}$ & $0.413^{+0.602}_{-0.610}$ \\
\hline
\end{tabular}
\caption{Current limits on the dark energy EoS}\label{tab:result}
\end{table}
One can see from this table that, given the current observation the
isocurvature perturbations  makes a change. However the
effect is small. As one can see from
Fig.\ref{fig:cmb}  the cause of the change is the isocurvature mode
which suppresses the power spectrum on large scale. Theoretically,
to have a sizable suppression, a smaller $w_e(a)$ is required in the
early universe. This explains why a smaller mean value of $w_e(a)$
is obtained in the mixed case given the current data with the
suppressed CMB quadrupole. In Fig.\ref{fig:w0wa} we plot the two
dimensional constraints on the dark energy EoS parameters $w_0$,
$w_a$. To show the importance of dark energy perturbations we  also
present the results obtained when the DE perturbations switched off
incorrectly. We can see from this plot that it brings  an  error
 $9\%$ on $w_0$ and $52.1\%$ on
$w_a$.

\begin{figure}[htbp]
\centering
\includegraphics[width=0.5\linewidth]{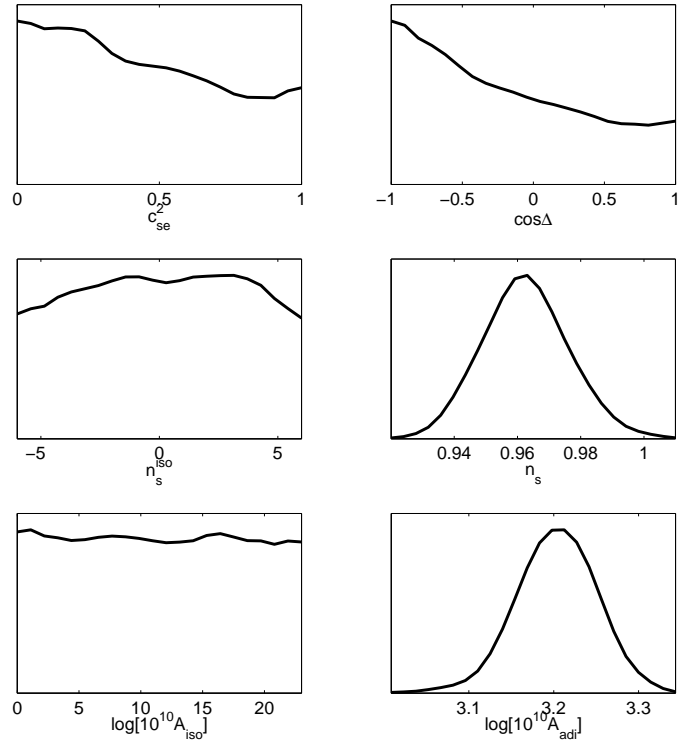}
\caption{The marginalized 1D probability distribution of the
cosmological parameters related to dark energy perturbation.
}\label{fig:cons}
\end{figure}

In Fig.\ref{fig:cons}, we plot the marginalized $1-D$ probability
distribution of the parameters related to the initial conditions.
The constraints on isocurvature parameters, such as $A_{\rm iso}$,
$n_{s}^{\rm iso}$ and $\cos\Delta$ of dark energy are weak. This is
understandable, since, as shown in Fig.\ref{fig:cmb} and
\ref{fig:lss}, the isocurvature perturbations of dark energy  mainly
make contribution at large scale, where the observational data are
limited. We note that for the mixed case, the mean value of the
adiabatic primordial perturbation amplitude $A_{\rm
adi}\sim2.5\times10^{-9}$ is slightly larger than in the adiabatic
case $A_{\rm adi}\sim2.4\times10^{-9}$, while the spectral index is
smaller. This indicates that, with respect to adiabatic
perturbation, the isocurvature mode has a negative effect on the
power spectra on large scale (small $k$), \ie, the mixed mode can
suppress the CMB TT angular power spectra at low values of $l$.
Moreover, the decrease of $\chi^2$,
$\Delta\chi^2=\chi^2_{adi}-\chi^2_{mix}\sim2$ gives a hint that the
mixed case is mildly favored by the data.

\section{Summary and Discussion}
\label{summary}

As is well known,  cosmological perturbations are crucially
important for understanding the CMB anisotropies and structure
formation. Up to now the theory of cosmological perturbation is very
successful and has been confirmed by the high precision observations
and experiments such as WMAP and SDSS. Since the perturbed spacetime
is determined by the perturbations of all of the matter components
in the universe, it is also important to study the dark energy
perturbation. If we naively switch off the dark energy perturbation,
the result would be misleading. Moreover, to be general, besides the
adiabatic perturbation which is mostly studied in the literature,
one should also consider the isocurvature perturbation. Because dark
energy couples very weakly to other matter it is not so easy to
construct a dark energy model which has purely adiabatic
perturbation. Dark energy isocurvature perturbations have important
application to lower the quadrupole  of the CMB angular power
spectrum  as needed by the COBE and WMAP
observations\cite{quadrupole}.

In this paper, we have studied in detail the effects of dark energy
isocurvature perturbations. We have included  dark energy
isocurvature perturbation in the  data analysis. By employing a
Markov Chain Monte Carlo method, we have performed a global analysis
of the determination of the cosmological parameters from current
astronomical observational data. We find that isocurvature
perturbations decease $\chi^2$ by about $2$, and has small effect on
other parameters. The current limit on the isocurvature initial
condition is weak.  We expect that future  precision measurements of
CMB and LSS on large angular scales, especially the measurements of
CMB-LSS cross correlations will lead to a tighter constraint.

\section*{Acknowledgements}
We thank Jun-Qing Xia, Yi-Fu Cai, Hong Li, Zuhui Fan, Charling Tao
and Hu Zhan for discussions. We thank Robert Brandenberger and
Taotao Qiu for helping us polish language. The calculation is
performed on Deepcomp7000 of Supercomputing Center, Computer Network
Information Center of Chinese Academy of Sciences. M.L. is supported
by the NSFC under Grants Nos. 11075074 and 11065004 and the
Specialized Research Fund for the Doctoral Program of Higher
Education (SRFDP) under Grant No. 20090091120054. X. Z. and J.L. are
supported in part by the National Natural Science Foundation of
China under Grants Nos. 10975142, 10821063 and by the 973 program
Nos. 1J2007CB81540002.

\end{document}